\documentclass[reprint,amsmath,amssymb,aps,pra,twocolumn,superscriptaddress]{revtex4-1}
\usepackage{graphicx}
\usepackage{dcolumn}
\usepackage{bm}
\usepackage{color}
\usepackage{setspace}

\begin{document}
\begin{spacing}{1}

\title{Dynamics of interacting bosons in a double well potential- \\ harmonic versus chirp modulation } 
\author{Sunayana Dutta}
\affiliation{Department of Physics, Indian Institute of Technology  Guwahati, Assam, India}
\author{Pankaj Kumar Mishra}
\affiliation{Department of Physics, Indian Institute of Technology  Guwahati, Assam, India}
\author{Budhaditya Chatterjee}
\affiliation{Department of Physics, Indian Institute of Technology  Kanpur, India}

\author{Saurabh Basu}
\affiliation{Department of Physics, Indian Institute of Technology  Guwahati, Assam, India}

\date{\today}

\begin{abstract}
We present effects of an external driving on the tunneling dynamics of interacting bosons confined in a double well potential. At large values of a periodic driving potential, the dynamics become chaotic, with a distinct difference between tuning of the amplitude 
or the phase of the driving term with regard to the route leading to chaos. For example, we find that a controlled increase in the amplitude 
of the driving term with a fixed phase leads to quasiperiodic route to  chaos. While tuning of the frequency with a fixed (large) amplitude 
leads to a crisis induced intermittency route to a chaotic dynamics, where the system intermittently visits different attractors in the 
phase space. Surprisingly, a chirp frequency superposed on a harmonic signal suppresses the chaotic motion, thereby yielding an orderly 
dynamics, pretty similar to a (damped) Rabi oscillation.
\end{abstract}
\maketitle

\section{Introduction}
Periodically driven systems, for example ultracold atoms in a periodically shaken optical lattice have received unprecedented attention following the realization of exotic behaviour of such systems. A 
paradigmatic example of these systems is the kicked top or kicked rotor model where a particle moving on a ring is subjected to 
periodic kicks \cite{Casati}. The display of transition from integrability to chaotic behaviour in certain limits \cite{Haake,Quantum,S} and dynamical Anderson localization \cite{Fishman,Reichl}, dynamical 
stabilization both in classical and quantum mechanics, coherent control of the phase transition from superfluid to Mott insulator \cite{Andre}, and parameter controlled adiabatic transformation of a static 
Bose-Einstein condensate into a dynamical Floquet condensate \cite{Heinisch} are few prominent features 
of these systems. Hence they facilitate investigation of a particularly rich research area for the classical chaos and quantum chaos communities.

The periodic perturbations can be used as a tensile experimental knob \cite{Ovadyahu,Iwai,Kaiser} to perceive synthetic matter.  For example, the opening of a gap in graphene by using light or controlled lattice phonons \cite{Oka,Kita} and the realization  of the Floquet topological insulator in a material that in the absence of driving yields a trivial topological insulator \cite{Linder,Wang}.

While a volume of literature exists in the field of periodically driven Bose Einstein condensates~\cite{Hai,Salmond,Xie}, however dynamical studies of a driven 
Bose Hubbard model have not received too much attention.
Motivated by such exciting possibilities, in the paper, we investigate the dynamics of a system of interacting bosons in a double well potential where the particles are interacting via a harmonic interaction potential. In an effort to go beyond periodic driving,  we have included the effect of a chirp frequency so as to compare and contrast with the case for periodic driving. A chirp is essentially a sinusoidal signal whose phase changes instantaneously at each time step and its correlation properties resemble an impulse function. In fact, adding chirp into the signal is an efficient tool in echo location systems, such as radars/sonars. In the low temperature physics experiment, chirp signal plays an important role in controlling the dynamics of atoms. Wright \textit{et. al} \cite{Wright} investigated the dynamics of collisions of ultracold Rb atoms induced by near-resonant frequency chirp light, where the rate of collisions can be controlled by  changing the pulse detunings.  In another experimental work \cite{Carini}, the effect of nanosecond-timescale chirp laser pulse on  the formation of $^{87}$Rb ultracold molecules is studied via photo-association experiments. It is observed that the chirping effect enhances the rate of formation of molecules.  However in our work, the harmonic and the chirp disturbances are considered as the agents that induce or suppress chaotic fluctuations in the double well dynamics of the bosons.

In the present work we introduce a general interaction comprising of a harmonic and chirp modulation denoted by, $c(t)=c_{0}\cos\phi(t)$, where $\phi(t)=\omega t+\beta t^2$ and further investigate the dynamics  of a system of bosons in the double well potential. The situation is identical with a two site Bose Hubbard model (BHM) \cite{Lu,Har,Lign,G,Eck,gati}, where, the interparticle interaction $c(t)$  strength is added to study the dynamical evolutions. At zero or low values of the strength of the interaction potential, the atoms execute Rabi oscillations. However the scenario changes to far from being Rabi-like, that is, chaotic as the interaction strength ($c_{0}$) is enhanced. Depending on the system  and forcing strength there can be  several routes to chaos, namely, quasiperiodicity \cite{bishop,Oteski,mishra},  period doubling \cite{period double}, intermittency \cite{intermittency}, crisis induced intermittency \cite{crisis,crisisInter}, and Ruelle-Takens-Newhouse~\cite{RTN}.    

In this paper we present different routes to chaos of the interacting bosons in presence of different natures of the time varying interaction potential. In general we notice two kinds of route to chaos depending on the nature of the forcing: (i) quasiperiodic route to chaos in the situation where, frequency of the harmonic modulation ($\omega$) is kept fixed at a particular value and the interparticle interaction strength is varied, and (ii) the crisis induced intermittency route to chaos for the system where the interaction strength is kept fixed, while  $\omega$ is varied within a certain range. 

The structure of the paper is as follows: Section II deals with the quantum dynamics of the system of bosons confined in a double well potential and described by a Bose Hubbard model (BHM) on a two site lattice. The dynamics is computed by mapping the model into one consisting of the components of 
the angular momentum (SU generators). In Sec. III, we discuss the dynamics of the condensate by computing the time evolution of the population 
between the two wells. Further, to support the behaviour of the time evolution of the system, the  power spectral density and phase space features of the population imbalance (occupation difference between left and right well) are analyzed. Finally we conclude our observations in Sec. IV.
\section{Equations of motion for the $SU(2)$ generators}

For the two mode approximation, the Bose Hubbard Hamiltonian for a system of $\textit{N}$ interacting bosons occupying the
weakly coupled low lying energy states can be written as \cite{S},
\begin{eqnarray}
\label{one}
\hat{H} &=& \frac{\epsilon}{2}(\hat{a}_{1}^{\dagger} \hat{a}_{2}+\hat{a}_{2}^{\dagger} \hat{a}_{1})
         +\frac{\gamma}{2}(\hat{a}_{1}^{\dagger} \hat{a}_{1}-\hat{a}_{2}^{\dagger}\hat{a}_{2})\nonumber\\
        & &-\frac{c(t)}{2}(\hat{a}_{1}^{\dagger}\hat{a}_{1}^{\dagger} \hat{a}_{1} \hat{a}_{1}
         +\hat{a}_{2}^{\dagger} \hat{a}_{2}^{\dagger}\hat{a_{2}} \hat{a_{2}}),
\end{eqnarray}
where, $\hat{a}_{1}^{\dagger} (\hat{a}_{2})$ are the creation (annihilation) operators
of bosons in the first (second) wells, $\epsilon$ is the coupling between the modes
(i.e., the tunneling parameter and $\epsilon < 0$ here), $\gamma$ is the energy difference between
the quantum states and \textit{c(t)} is the driving term that disturbs the system which is assumed to be time-dependent in our case (elaborated below).
For $c(t)=c_{0}$, we recover the familiar BHM. 
The time-dependent interaction term has been studied earlier and periodic modulations for some 
or each of the terms in Eq.~\eqref{one} have received some attention as well \cite{H,G,W}.
In contrast to the above, we consider a chirp modulation in the driving term, in addition to a periodic component which we denote by, 
\begin{eqnarray}
\label{two}
c(t) &=& c_{0}\cos(\omega t+\beta t^{2}), 
\end{eqnarray}
where, $c_{0}$ is the amplitude of the disturbance, $\omega$ and $\beta$ denote harmonic and chirp modulations, respectively.
Here, both the energy scales, namely $\gamma$ and $c_{0}$ are measured in the units of $\epsilon$.

Clearly, the Hamiltonian in Eq.~\eqref{one} commutes with the total number of particles 
$N (= \hat{a}_{1}^{\dag} \hat{a}_{1}+ \hat{a}_{2}^{\dag} \hat{a}_{2})$ and hence \textit{N} 
is conserved. Further the Hamiltonian can be written in a symmetric fashion (upto a constant term depending on \textit{N}) of the form,
\begin{eqnarray}
\label{three}
\hat{H} &=& \frac{\epsilon}{2}(\hat{a}_{1}^{\dag} \hat{a}_{2}+\hat{a}_{2}^{\dag} \hat{a}_{1})+\frac{\gamma}{2}(\hat{a}_{1}^{\dag} \hat{a}_{1}
              -\hat{a}_{2}^{\dag}\hat{a}_{2})-\nonumber\\
        & &      \frac{c(t)}{4}(\hat{a}_{1}^{\dag} \hat{a}_{1}-\hat{a}_{2}^{\dag} \hat{a}_{2})^{2}.
\end{eqnarray}
 \begin{figure}[!t]
\hspace*{-0.3cm}
\includegraphics[scale=0.60]{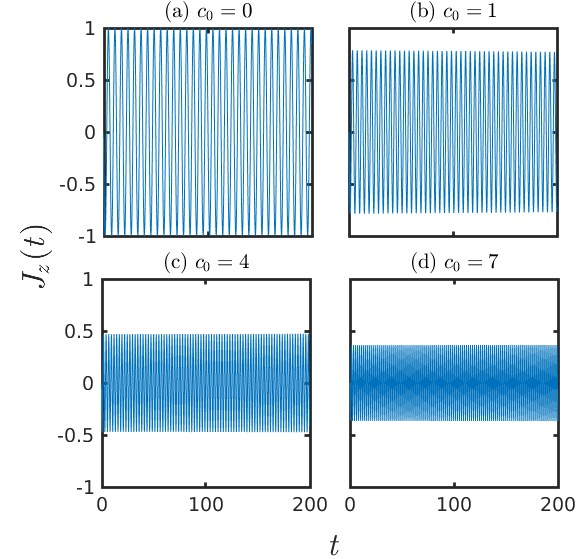}
\caption{Time evolution of the population imbalance $J_{z}(t)$ for the static interaction ($\omega=0$ and $\beta=0$).
(a) corresponds to $c_{0}=0$, (b) corresponds to $c_{0}=1$, (c) corresponds to $c_{0}=4$ and (d) corresponds to $c_{0}=7$. Here $c_{0}$ values are scaled by the number of bosons, $N$ and $t$ is in units 
of inverse of the energy scale, $\epsilon$.
\label{Fig:1}}
\end{figure}
\begin{figure*}[htbp]
\hspace*{-1cm}
\includegraphics[scale=0.60]{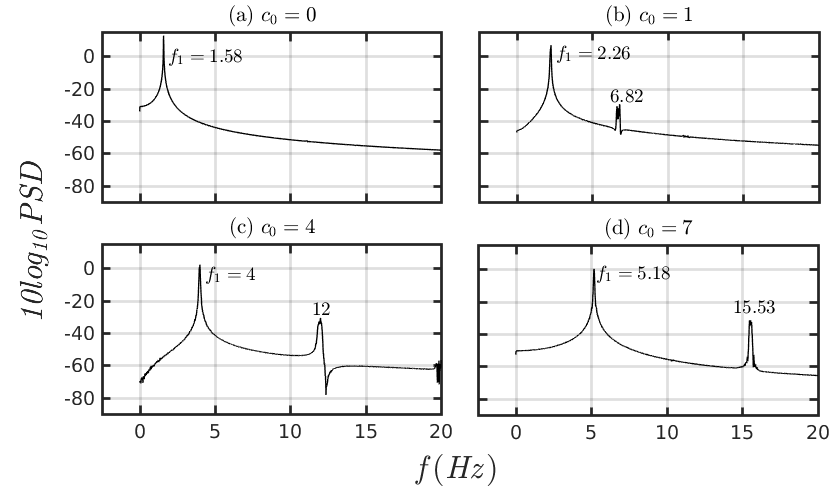}
\caption{Power spectral density, $PSD$ for the static interaction ($\omega=0$ and $\beta=0$) and different values of $c_{0}$, namely (a) corresponds to $c_{0}=0$, 
(b) corresponds to $c_{0}=1$, (c) corresponds to $c_{0}=4$, and (d) corresponds to $c_{0}=7$.
 Additional frequencies appearing from (b)-(d) are simply the harmonics of the fundamental frequency, $f_{1}$. 
\label{Fig:2}}
\end{figure*} 
To solve Eq.~\eqref{three}, it is convenient to introduce the SU(2) generators, namely the components of the angular momentum, 
$J_{x}$, $J_{y}$ and $J_{z}$ as,
\begin{eqnarray}
\label{four}
J_{x}  &=& \frac{1}{2}(\hat{a}_{1}^{\dag} \hat{a}_{2}+\hat{a}_{2}^{\dag} \hat{a}_{1}) ~;\nonumber\\ 
J_{y}  &=& \frac{1}{2i}(\hat{a}_{1}^{\dag} \hat{a}_{2}-\hat{a}_{2}^{\dag} \hat{a}_{1}) ~;\\
J_{z}  &=& \frac{1}{2}(\hat{a}_{1}^{\dag} \hat{a}_{1}-\hat{a}_{2}^{\dag} \hat{a}_{2}),\nonumber
\end{eqnarray}
where, the angular momentum operators, $J_{i}$ obey the usual commutation relation, $[J_{i},J_{j}] =
i J_{k}$ ($\hbar=1$). With the above assumptions the Hamiltonian assumes the form,
\begin{eqnarray}
\label{five}
\hat{H} &=& \epsilon J_{x} + \gamma J_{z} - c(t) J_{z} ^{2}
\end{eqnarray}
The analogy between Eqs.~\eqref{three} and ~\eqref{five} can be made clearer with the help of an eigenstate, $|\chi _{n}\rangle$ 
which denotes \textit{n} particles in the first well and $(N-n)$ in the second one.
The operators $J_{x}$, $J_{y}$ and $J_{z}$ are defined by their action on $|\chi _{n}\rangle$ which are stated here for the sake of completeness
as,
\begin{eqnarray}
\label{six}
J_{x}^{2}|\chi _{n}\rangle &=& \frac{1}{4}[\sqrt{(n+1)(n+2)(N-n)(N-n-1)} |\chi _{n+2}\rangle \nonumber\\
                           & &+(n(N-n+1)+(n+1)(N-n)) |\chi _{n}\rangle \nonumber \\
                           & & +\sqrt{n(n-1)(N-n+1)(N-n+2)} |\chi _{n-2}\rangle]; \nonumber
\end{eqnarray}
\begin{eqnarray}
\label{seven}
J_{y}^{2}|\chi _{n}\rangle &=& \frac{1}{4}[-\sqrt{(n+1)(n+2)(N-n)(N-n-1)}|\chi _{n+2}\rangle \nonumber\\
                           & & + (n(N-n+1)+(n+1)(N-n))|\chi _{n}\rangle \\
                           & & - \sqrt{n(n-1)(N-n+1)(N-n+2)}|\chi _{n-2}\rangle];\nonumber\\ 
J_{z}^{2}|\chi _{n}\rangle &=& \frac{1}{4}[n^{2}-2n(N-n)+(N-n)^2]|\chi _{n}\rangle; \nonumber  
\end{eqnarray}
Thus for the total angular momentum, $\vec{J}^{2} (={J_{x}}^{2}+{J_{y}}^{2}+{J_{z}}^{2})$, one can write,
\begin{eqnarray}
\label{eight}
\vec{J}^{2}|\chi _{n}\rangle &=& \frac{N}{2}\Big(\frac{N}{2}+1\Big)|\chi _{n}\rangle;
\end{eqnarray}
Comparing with the eigenvalue $j(j+1)$ for ${\vec{J}}^{2}$, we obtain $j={\frac{N}{2}}$.

Thus the components of the angular momentum vector can be described on a sphere (called a Bloch sphere) in three dimensions.
The \textit{z}-component, that is $J_{z}$ denotes the difference in population density between
the two wells.
 \begin{figure*}[htbp]
\centering
\includegraphics[scale=0.60]{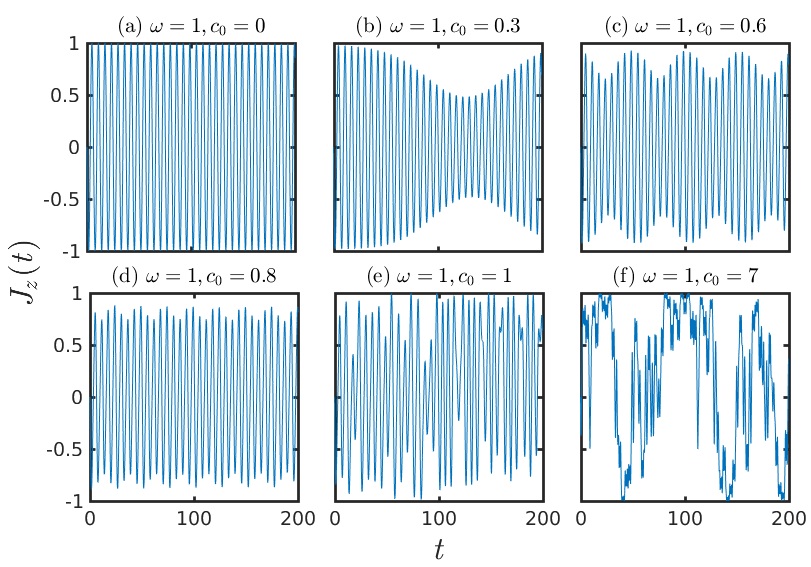}
\caption{Time evolution of the population imbalance, $J_{z}(t)$ in presence of harmonic interaction for $\omega=1$. 
(a) Presence of Rabi oscillations for $c_{0}=0$. (b) Appearance of two frequencies $f_1$ and $f_2$ at $c_{0}=0.3$. For (c) and (d), the observed 
dual periodicity becomes prominent for $c_{0}=0.6$ and $c_{0}=0.8$ respectively. Aperiodic nature is observed with further increase in 
$c_{0}$ in (e) and (f). Here $c_{0}$ values are scaled by the number of bosons, $N$ and $t$ is in units 
of inverse of the energy scale, $\epsilon$.
\label{Fig:3}}
\end{figure*}

The dynamics of the system is obtained by computing the equation of motion (EOM),
\begin{eqnarray}
\label{nine}
\dot{J}_{i} &=& \frac{1}{i} [J_{i},H].
\end{eqnarray}
This EOM yields three coupled equations of the form,
\begin{eqnarray}
\label{ten}
\dot{J}_{x} &=& -\gamma J_{y}+c(t)(J_{y}J_{z}+J_{z}J_{y})\nonumber\\
\dot{J}_{y} &=& \gamma J_{x}-\epsilon J_{z}-c(t)(J_{x}J_{z}+J_{z}J_{x}) \\
\dot{J}_{z} &=& \epsilon J_{y} \nonumber
\end{eqnarray}

These coupled set of equations are now solved in presence of the interaction potential given by, $c(t)$ (see Eq.~\eqref{two}) to obtain the time evolved trajectories 
of $J_{x}$, $J_{y}$ and $J_{z}$ which are the components of the Bloch vector. We have considered the following initial condition for ($J_{x}$, $J_{y}$, $J_{z}$), 
namely (0,1,0) at $t=0$, which implies both the wells contain equal number of bosons. Hence the tunneling dynamics are obtained by solving Eq.~\eqref{ten}.

In the following section we will present our analysis of the tunneling dynamics of the interacting Bose systems in the double well potential for specific choices of $c(t)$.
\section{Results}

In order to have an intuitive idea about the dynamical behaviour of the system in a double well potential, we systematically investigate the time evolution of population imbalance between the wells, $J_{z}(t)$ 
(as given in Eq.~\eqref{four}). 
To get a deeper understanding of the trajectory of the system and onset of chaotic behaviour therein, the power spectral density, $PSD$ is analyzed which is defined as,
\begin{eqnarray}
\label{eleven}
PSD &=& \frac{1}{2\pi \mathcal{N}}|J_{z}(\mathcal{N},f,\tau)|^{2},
\end{eqnarray}
where $J_{z}(\mathcal{N},f,\tau)$ is the discrete Fourier transform of the population imbalance, $J_{z}(t)$ evaluated at $t=k\tau$ ($k=0,1,...\mathcal{N}$ and $\mathcal{N}$ is the 
length of the discrete time series).
Further the corresponding phase space projection of $J_{z}(t+\tau)$ and $J_{z}(t)$
 are plotted to distinguish different kinds of dynamics as the function of the driving term, $c(t)$ that emerge in a double well potential.
 
\subsection{Static interaction, $c(t)=c_{0}$}
First we consider the case of static interaction without presence of any time variation in $c(t)$. In absence of interaction, $c_{0}=0$, 
usual Rabi oscillation is observed (Fig.~\ref{Fig:1}(a)) where the particles periodically move back and forth among the two wells. As the interaction strength, $c_{0}$ is gradually increased one observes damped 
Rabi oscillations, that is evident from Fig.~\ref{Fig:1}(b)-(d) where the amplitudes of $J_{z}(t)$ diminish indicating that only a fraction of the atoms move from one well to another.

 In Fig.~\ref{Fig:2} we show the power spectral density ($PSD$) of the temporal evolution 
 of the population dynamics. We find that without any interaction, the fundamental frequency of the system appears at 
 $f_{1}\simeq1.6$, 
 which is benchmarked as the frequency of Rabi oscillations. 
 
 With the inclusion of the interaction term, 
 the fundamental frequency gets shifted to larger values due to the change in the period of oscillations of the system 
 (damped Rabi oscillations). For example, in Fig.~\ref{Fig:2}(b) $f_{1}$ gets shifted to $2.26$ for $c_{0}=1$. In addition, we observe that for 
 non-zero values of $c_{0}$, the power spectral density exhibits two peaks instead of one. However, a closer look  
 reveals that the second frequency present in the plot at $6.82$, which is  simply a harmonic (or an overtone) of 
 the fundamental frequency of the system, that is $f_{2}\simeq6.82\simeq3f_{1}$.  Similarly in Figs.~\ref{Fig:2}(c) and ~\ref{Fig:2}(d), respectively for $c_{0}=4$ and $c_{0}=7$,
 two peaks are observed at $f_{1}=4$  and $f_{2}=12$ and $f_{1}=5.18$ and $f_{2}=15.53$. Again they satisfy $f_{2}\simeq3f_{1}$. 
 Thus it is clear that upon increasing (only) the strength of the interaction potential (and no time variation yet), 
 the dynamics of bosons in a double well involve only one frequency as it should be. This is also apparent from the dynamics presented in Fig.~\ref{Fig:1}.
\begin{figure*}[htbp]
\hspace*{-1.1cm}
\includegraphics[scale=0.60]{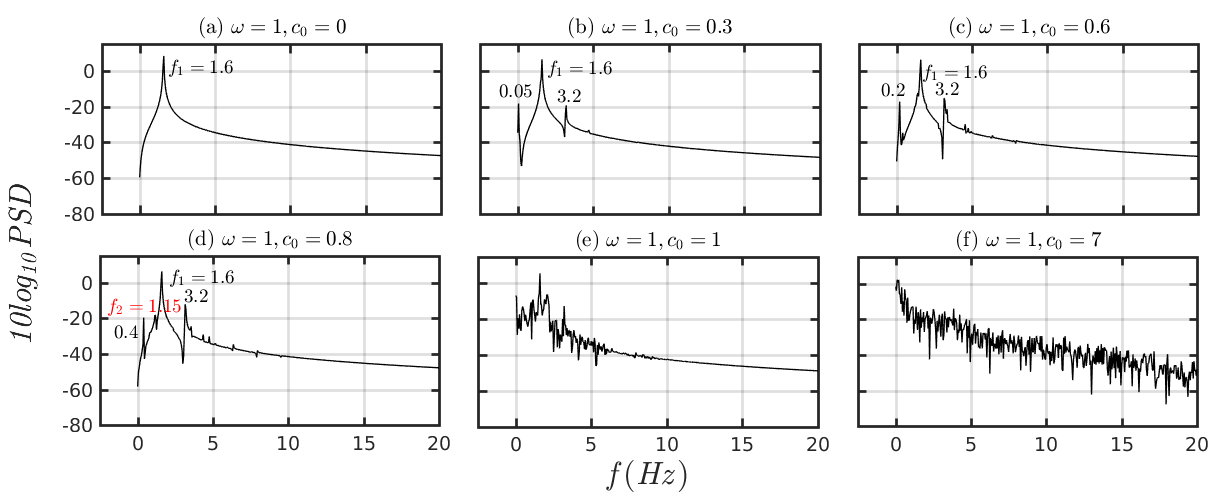}

\caption{ Power spectral density, $PSD$ in presence of the harmonic interaction for $\omega=1$ and 
for different values of $c_{0}$, namely (a) $c_{0}=0$, (b) $c_{0}=0.3$, (c) $c_{0}=0.6$, (d) $c_{0}=0.8$, (e) $c_{0}=1$, and (f) $c_{0}=7$.
Upto $c_{0}=0.6$, additional frequencies appearing are harmonics of the frequency $f_{1}$. However from $c_{0}=0.8$ onwards more frequencies 
start populating the spectrum. At $c_{0}=7$, it becomes heavily populated, alongwith an exponential decay signaling onset of chaos.
\label{Fig:4}}

\end{figure*}

\begin{figure}[!t]
\hspace*{-1.8cm}
\includegraphics[scale=0.60]{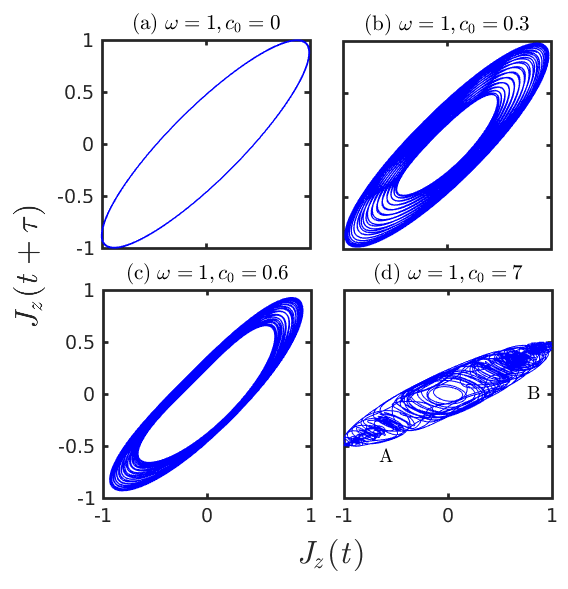}
\caption{Phase space projection of $J_{z}(t)$ for harmonic modulation (constant $\omega$ and varying $c_{0}$).
(a) periodic regime for $c_{0}=0$ shows an ellipse, (b)-(c) quasiperiodic regime for $c_{0}=0.3$ and $c_{0}=0.6$ showing a torus and (d) chaotic regime
for $c_{0}=7$. Chaotic trajectory is confirmed from the presence of two attractors marked by A and B.
\label{Fig:5}}
\end{figure}
\subsection{Harmonic interaction}
Now we include the harmonically driven term in the interparticle interaction to see the effect on the dynamics of the system (that is, $\omega\ne0$ and $\beta=0$ in Eq.~\eqref{two}), such that $c(t)=c_{0}\cos(\omega t)$. Let us further divide this scenario into two cases and explore the dynamics : first by keeping $\omega$ constant (say at $\omega=1$) and by varying $c_{0}$, and in the other, keeping $c_{0}$ fixed at a certain value and varying $\omega$.
\vskip 0.1 in
{(i) First Case ($\omega=1$ and $c_{0}$ is varied):}
\vskip 0.05 in
In this case, with a gradual increase of the interaction strength, $c_{0}$, it is observed that for $c_{0}=0.3$, the time evolved
population imbalance, $J_{z}(t)$, shows appearance of a dual periodicity (Fig.~\ref{Fig:3}(b)), that is, both slow and fast oscillations are 
present, a hallmark signature of periodically driven systems. One of the periodicities corresponds to the natural dynamics with
the other denoting the driving frequency.
The slow oscillations become more frequent 
with increase in values for $c_{0}$ (see Fig.~\ref{Fig:3}(c)and ~\ref{Fig:3}(d)). However
the observed dual periodicity in the behaviour of $J_{z}(t)$ gradually disappears at large values of $c_{0}$. 
For very high values of $c_{0}$ ($c_{0}=7$), it is observed that the behaviour of $J_{z}(t)$ becomes completely aperiodic in nature 
(Fig.~\ref{Fig:3}(f)). There is an onset of chaotic dynamics in the system. Hence $c_{0}$ is fixed at this value, namely $c_{0}=7$ for subsequent discussion.
\begin{figure*}[htbp]
\centering
\includegraphics[scale=0.60]{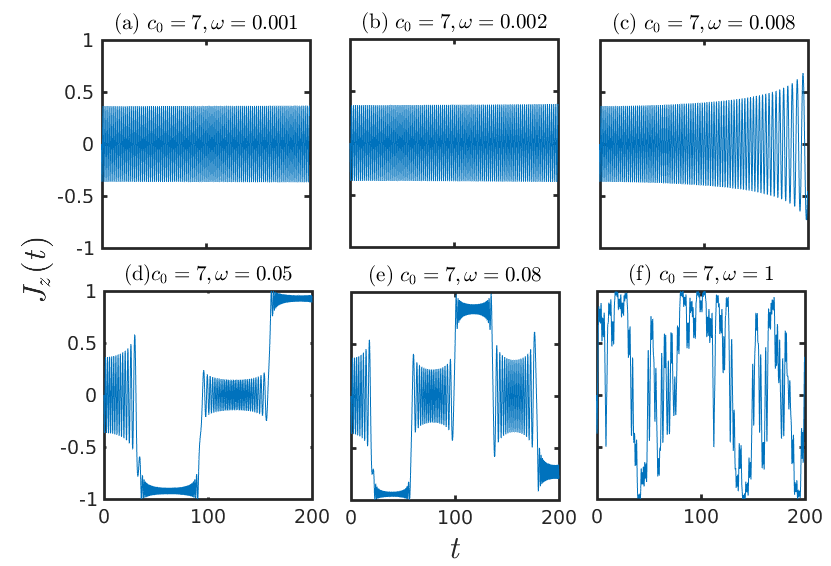}
\caption{ Time evolution of population imbalance, $J_{z}(t)$ in presence of harmonic modulation with constant $c_{0}$, namely $c_0=7$.
$\omega$ is varied, namely (a) $\omega=0.001$, (b) $\omega=0.002$, (c) $\omega=0.008$, (d) $\omega=0.05$, (e) $\omega=0.08$, and (f) 
$\omega=1$. Here $c_{0}$ values are scaled by the number of bosons, $N$ and $t$ is in units 
of inverse of the energy scale, $\epsilon$.
\label{Fig:6}}
\end{figure*}

To probe deeper into the above scenario, we compute the $PSD$ and systematically analyze the presence of different frequencies therein. In Fig.~\ref{Fig:4} we show the $PSD$ for the case when the frequency is fixed ($\omega=1$) and $c_0$ is varied in the range that is from  $c_0=0$ to $c_0=7$. Without interaction, the $PSD$ peaks at the fundamental frequency $f_{1}=1.6$ (Fig.~\ref{Fig:2}(a) and ~\ref{Fig:4}(a)). For small and finite values of the interaction strength (for example, $c_{0}=0.3$), two 
peaks in the $PSD$ appears apart from $f_{1}$. One at $f\simeq1.6$ and another at $f\simeq0.05$ (marked in Fig.~\ref{Fig:4}(b)). These two frequencies are simply the harmonics of $f_1$. Upon further increase of $c_0$, that is to $c_0=0.8$ results the appearance of the peaks in the $PSD$ at frequencies $f_2=1.15$ and $f_1=1.6$. Since $f_{1}$ and  $f_{2}$ are independent, this indicates the involvement of two frequencies of oscillations in the time evolution of $J_{z}(t)$, which is a signature of the quasiperiodic nature of the system. With further increase in the interaction strength, we find that more frequencies near $f_1$ and $f_2$ start getting populated. Finally, for very large interaction strength, that is, $c_{0}=7$, the $PSD$ shows an exponential decay with fully populated pattern over the entire frequency range (Fig.~\ref{Fig:4}(f)). This is the distinguishing feature of the onset of chaos. 
Thus for $c_{0}=7$, the observed aperiodic time evolution of $J_{z}(t)$ displays chaotic behaviour for the tunneling dynamics (see Fig.~\ref{Fig:3}(f)).
Hence we obtain a series of transitions, namely, from periodic to quasiperiodic and finally to chaotic dynamics as the strength of the harmonic interaction is enhanced. Thus fixed $\omega$  and varying $c_{0}$ case signals towards a {\it quasiperiodic route to chaos}.

Next we plot the phase space projection defined by $J_{z}(t+\tau)$ vs $J_{z}(t)$ to establish the observed chaotic behaviour in the system.
Without any interaction ($c_0=0$), the phase space projection shows an usual elliptical trajectory (Rabi) depicting periodic behaviour of the system (Fig.~\ref{Fig:5}(a)).
While for the quasiperiodic regime (small, but finite $c_{0}$), the elliptical trajectory is observed with distinct width (see Fig.~\ref{Fig:5}(b)-~\ref{Fig:5}(c)) denoting a quasiperiodic behaviour.
Finally, for $c_{0}=7$, the phase space trajectory traces out lines around two attractors that depict onset of a chaotic behaviour (Fig.~\ref{Fig:5}(d)).
The two attractors, although not very distinct, are located in the vicinity of the points A and B as marked in Fig.~\ref{Fig:5}(d).
Thus the phase plots corroborate the information obtained from the spectral densities.
\begin{figure*}[htbp]
\hspace*{-1.5cm}
\includegraphics[scale=0.60]{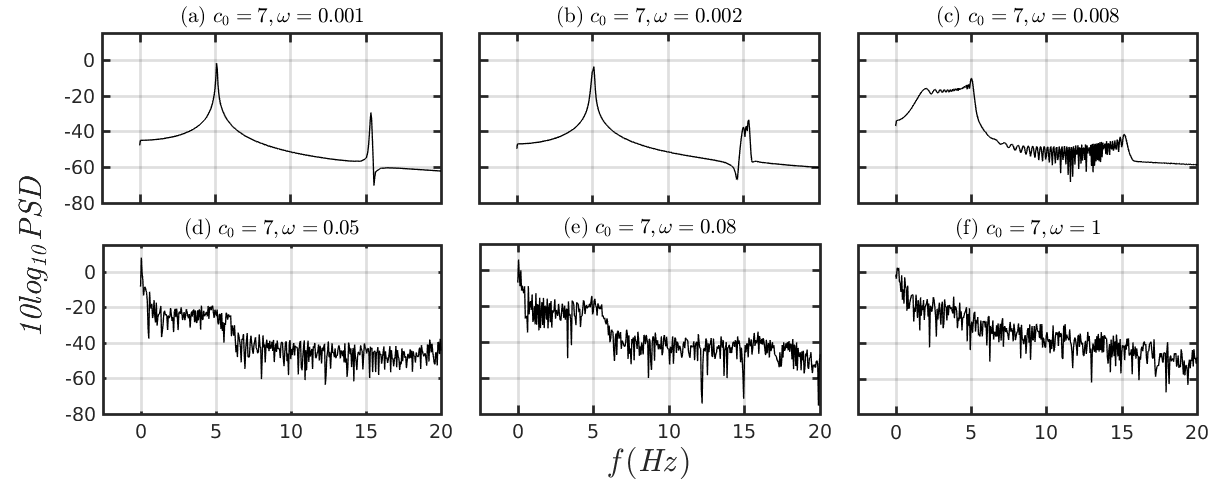}
\caption{Power spectral density $PSD$ for a constant $c_0$, $c_{0}=7$ and varying $\omega$.
(a) and (b) correspond to periodic state for $\omega=0.001$ and $\omega=0.002$. (c)-(e) correspond to crisis induced intermittency at $\omega=0.008$, $\omega=0.05$, and $\omega=0.08$. Finally (f) corresponds to chaotic state at $\omega=1$.  
\label{Fig:7}}
\end{figure*}
\vskip 0.1 in 
(ii) Second Case ($c_{0}=7$ and $\omega$ is varied):
\vskip 0.05 in
Now to see the effect of the harmonic driving alone, $c_{0}$ is fixed at $7$ and subsequently  $\omega$ is varied between 0 to 1. In Fig.~\ref{Fig:6} we show the temporal evolution of the population imbalance, 
$J_{z}(t)$ in this case. For very small values of $\omega$ (say $\omega=0.0001$, $0.002$), we find  that initially $J_{z}(t)$ exhibits damped Rabi oscillations with a single frequency (Fig.~\ref{Fig:6}(a) and (b)). On further increase  in $\omega$,  say for $\omega=0.008$ onwards, we observe appearance of multiple intermediate states in $J_{z}(t)$. The system fluctuates about the intermediate states for some time before making transition to another states.
The switching between the states happens to be intermittently (Fig.~\ref{Fig:6}(c)). This feature becomes more distinct with further increase in $\omega$ and occurs for smaller time spans (see Fig.~\ref{Fig:6}(d) and ~\ref{Fig:6}(e)). Finally at 
$\omega=1$, the dynamics of $J_{z}(t)$ become completely aperiodic (see Fig.~\ref{Fig:6}(f)) as we have noted earlier. However the route to chaos appears to be different in this case. In what follows we will systematically analyze the appearance of the chaos in the system by analyzing the $PSD$ for this case.
\begin{figure}[!t]
\hspace*{-1.8cm}
\includegraphics[scale=0.60]{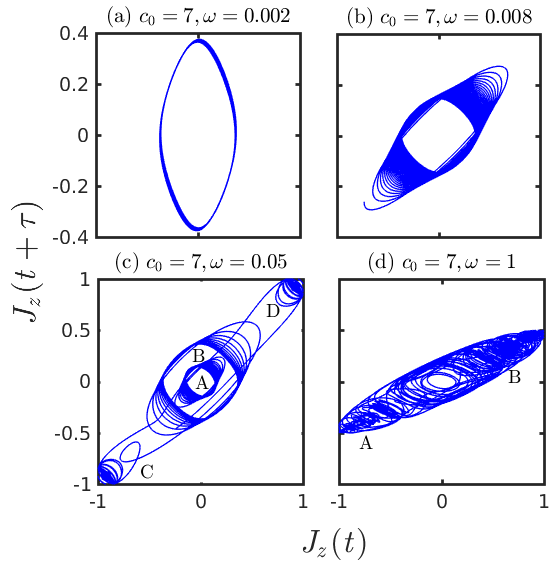}
\caption{Phase space projection of $J_{z}(t)$ for harmonic modulation ($c_{0}=7$ and varying $\omega$).
(a) $\omega=0.002$, (b) $\omega=0.008$, (c) $\omega=0.05$ and (d) $\omega=1$. The orbit is regular for $\omega=0.002$ which turns into torus (quasiperiodic) at $\omega=0.008$. At $\omega=0.05$ the system keeps on hopping intermittently between four attractors. At $\omega=1$ orbit turns out to be completely chaotic. 
\label{Fig:8}}
\end{figure}
\begin{figure}[htbp]
\hspace*{-0.9cm}
\includegraphics[scale=0.60]{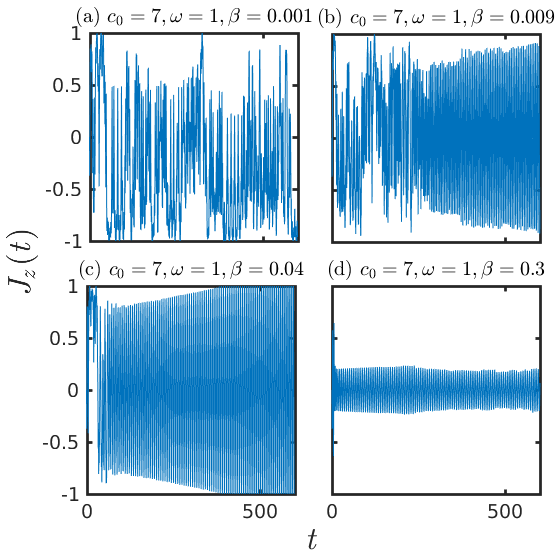}
\caption{Time evolution of population imbalance, $J_{z}(t)$ in presence both harmonic and chirp modulation. Here we keep $\omega=1$ and varying 
$\beta$ gradually for $c_{0}=7$. Transition from aperiodic to periodic trajectory is observed with gradual increase in $\beta$.
\label{Fig:9}}
\end{figure}
\begin{figure*}[!t]
\centering
\includegraphics[scale=0.60]{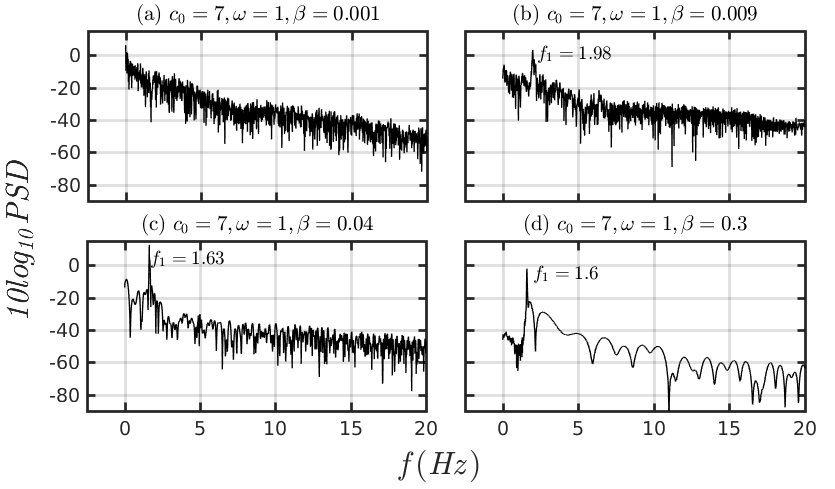}
\caption{Power spectral density $PSD$ in presence both harmonic ($\omega$) and chirp modulation ($\beta$). 
The frequency of harmonic modulation and amplitude are kept fixed at $\omega=1$ and $c_{0}=7$. (a) corresponds to $\beta=0.001$, (b) corresponds to $\beta=0.009$, (c) corresponds to $\beta=0.04$,
and (d) corresponds to $\beta=0.3$. With increase in $\beta$, the system restores its periodic behaviour. At $\beta=0.3$ the $PSD$ is peaked at the frequency $f_1=1.6$ which is similar to the Rabi frequency. 
\label{Fig:10}}
\end{figure*}

The scenario of approaching to a chaotic trajectory on varying $\omega$ becomes clearer  as we analyze  the $PSD$. In Fig.~\ref{Fig:7} we show the $PSD$ for $c_0=7$ and varying the frequency in the range between $\omega=0.002$ to $\omega=1$. We find that the $PSD$ gets populated in a random fashion from $\omega=0.002$ onwards (Fig.~\ref{Fig:7}(b)) 
and this random occurrence of a large number of new frequencies becomes more prominent with further increase in 
$\omega$. Finally for $\omega=1$ it leads to a chaotic behaviour which is quite evident from the exponential decay of the $PSD$ accompanied by the presence of large number of frequencies (Fig.~\ref{Fig:7}(d)). The transition in the dynamics from regular to a chaos happens due  to the {\textit{crisis induced intermittency route to chaos}}.

Hence with both the cases described above advocate chaotic scenario at $c_{0}=7$ and $\omega=1$, however the routes to this transition 
to a chaotic regime are clearly different. To summarize, the first case suggests of a quasiperiodic route to chaos, while the latter bears fingerprints 
of an intermittency route to chaos. These are among the central results of our paper, where a controlled variation of different parameters of a  
driven system renders different pathways leading to chaos.

A more lucid understanding of the above scenario can be offered in the following sense. Ramping up the amplitude ($c_0$) of the driving term 
leaving the angular frequency $(\omega)$ all the while to have a constant value indicates a systematic way of perturbing the system. The dual, 
that is, slow and fast oscillations are preserved. Thus the chaos induced in this case is distinct from the one in which the amplitude is kept fixed and 
$\omega$ is varied. The later has different ramifications as the frequency associated with the driving term is constantly varied, however the 
natural frequency of the tunneling from one well to another remains the same. At large driving frequencies (here $\omega=1$), a chaotic 
dynamics emerges which is distinct than the one noted in the previous case.

Finally, we study the phase space projection of the population imbalance, $J_{z}(t)$. For small value of $\omega$ ($\omega=0.001$), it is observed that 
the phase space trajectory traces out the usual elliptical orbit (Fig.~\ref{Fig:8}(a)) as the dynamics remain periodic at this value of $\omega$.  The regular orbits turns into torus (quasiperiodic) at $\omega=0.008$ as shown in the Fig.~\ref{Fig:8}(b). In Fig.~\ref{Fig:8}(c) we find that for  $\omega=0.05$, the system intermittently keeps switching between the four torus (marked as A, B, C, D) which is also termed as crisis induced intermittency~\cite{crisisInter}. At large frequency, namely, $\omega=1$ the orbits turns out to be completely chaotic. Here it appears that the system remains around one of the orbit (torus) for sometime and further makes transitions to another in an intermittent way. This frequent visits to the different intermediate states lead to the chaotic behaviour in the system.

\begin{figure}[htbp]
\hspace*{-0.9cm}
\includegraphics[scale=0.60]{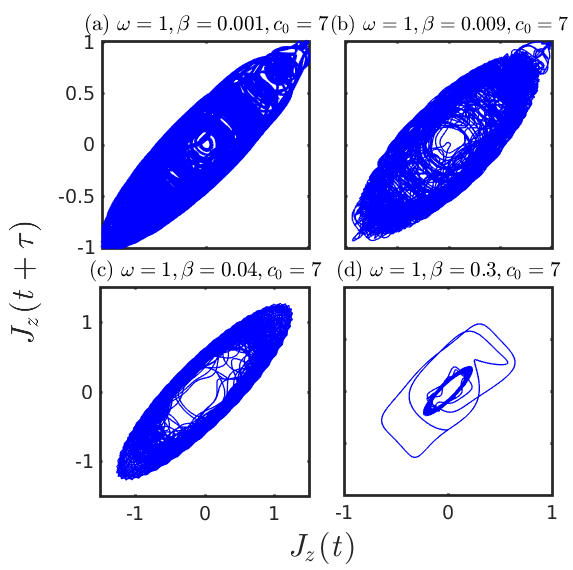}
\caption{Phase space projection of $J_{z}(t)$ in presence of both harmonic and chirp modulation with $c_{0}=7$, $\omega=1$ 
and varying $\beta$.
(a) corresponds to $\beta=0.001$, (b) corresponds to $\beta=0.009$, (c) corresponds to $\beta=0.04$ and 
(d) corresponds to $\beta=0.3$. The orbit is completely chaotic for $\beta=0.001$ which turns into a periodic one at $\beta=0.3$.
\label{Fig:11}}
\end{figure}
\subsection{A chirp modulation ($\beta\ne 0$)}
Let us superpose a chirp signal on the harmonic interparticle interaction and investigate what effects does it bring on the tunneling dynamics. We continue with $c_{0}=7$ and $\omega=1$. Now we vary $\beta$ (in units of square of the energy scale) (see  Eq.~\ref{two}) within a certain range. For very small value of $\beta$ (namely, $\beta=0.001$), the time evolution of the population imbalance, $J_{z}(t)$, shows aperiodic motion that is chaotic in nature (similar to Fig.~\ref{Fig:3}(f)). As we gradually increase 
$\beta$, periodicity in the evolution of $J_z(t)$ is restored partially as seen in Figs.~\ref{Fig:9}(b) and ~\ref{Fig:9}(c) for $\beta=0.009$ and 0.04, respectively. With further increase in $\beta$, for example, $\beta\sim0.3$, the dynamics of $J{z}$ becomes periodic in nature (shows damped Rabi oscillation similar to the case shown in Fig.~\ref{Fig:1}(b)-(d)), which is clear from Fig.~\ref{Fig:9}(d). 

The above feature suggests that addition of a small amount of chirp signal in the forcing term suppresses the chaotic fluctuations in the tunneling dynamics where the latter was generated by the harmonic term in the driving.

The $PSD$ yields deeper insights into the observed aperiodic-to-periodic transition in presence of a chirp modulation. For $\beta=0.001$, the $PSD$ shows exponential decay with a large number of frequencies being present, which culminates into a  chaotic behaviour (Fig.~\ref{Fig:10}(a)). With further increase in $\beta$, say for $\beta=0.009$, development of one prominent peak is observed at a frequency $f_{1}=1.98$, along with some fluctuations hinting towards the emergence of an orderly motion. At $\beta=0.04$, the fluctuations gradually subside, and a prominent frequency shifts to a value $f_{1}\sim1.6$. It may be noted that this is 
the same frequency observed corresponding to the case of no interaction (similar to Fig.~\ref{Fig:2}(a)). Finally, for 
$\beta=0.3$, the power spectral density shows a distinct peak precisely at $f_{1}=1.6$ that indicates the dynamics to be orderly in nature that 
is the atoms execute (damped) Rabi-like oscillations. Thus the chirp modulation brings back periodic nature to the tunneling dynamics by suppressing the observed chaotic behaviour at large interaction  strengths.  

A physical feel for the above result can be obtained in the following sense. Since the chirp term involves $\beta t^{2}$, 
it oscillates faster than the harmonic term  and hence the system feels as if being subjected to a constant force owing to 
the fact that the chirp oscillations are significantly  faster than the natural frequency of the motion of the particles between the two wells.

To confirm the re-entrant of periodic behaviour, we also study the phase space projection of the population imbalance $J_{z}(t)$. It is observed that for small value of $\beta$ ($\beta=0.001$), 
the phase space orbit appears to be completely chaotic in nature as shown in Fig.~\ref{Fig:11}(a). 
However the orbit takes 
the shape of a toroid for $\beta=0.04$ (Fig.~\ref{Fig:11}(c)). Finally, further increase in $\beta$ ($\beta=0.4$) leads to emergence 
of almost periodic
orbit rendering to the onset of a orderly motion as shown in Fig.~\ref{Fig:11}(d).

\section{Conclusions}
We have investigated the tunneling dynamics of a system of bosons in a double well potential in presence of 
a harmonic interaction potential. At large values of the interaction strength, the system makes a transition to a chaotic state. This feature is complemented very well by investigating  the power spectral density and phase space projection. In particular, 
for the case of harmonic interaction potential, different routes to chaos emerge depending on whether the amplitude or the phase 
of the interaction term being tuned. 
In the former case, we observed that the route to chaos is quasiperiodic, while, for the latter case, the the route to chaos is 
through the crisis induced  intermittency. Finally we have obtained  that superposition of a chirp modulation to the driving term restores periodicity in 
the tunneling dynamics of bosons in a double well potential.

\end{spacing}

\begin{thebibliography}{9}
\bibitem{Casati} 
{\it {Stochastic Behavior of a Quantum Pendulum under a
Periodic Perturbation, in Stochastic Behavior in Classical
and Quantum Hamiltonian Systems}}, Lecture Notes in
Physics, Vol. 93, G. Casati, B. V. Chirikov, F. M. Izraelev, and J. Ford, edited by Giulio Casati and Joseph Ford, 
Berlin Heidelberg; Springer (1979).
\bibitem{Haake} F. Haake, M. Kus, and R. Scharf, Z. Phys. B {\bf{65}}, 381 (1986).
\bibitem{Quantum} {\it{Quantum Signatures of Chaos}}, F. Haake, Berlin; Springer (2001).
\bibitem{S} M. P. Strzys, E. M. Graefe, and H. J. Korsch, New Journal of Physics {\bf{10}}, 013024 (2008).
\bibitem{Fishman} S. Fishman, D. R. Grempel, and R.E. Prange, Phys. Rev. Lett. {\bf49}, 509 (1982).
\bibitem{Reichl} {\it The Transition to Chaos: Conservative Classical Systems and Quantum Manifestations}, L. E. Reichl, New York; Springer (2004). 
\bibitem{Andre} A. Eclardt, C. Weiss, and M. Holthaus, Phys. Rev. Lett. {\bf 95}, 260404 (2005).
\bibitem{Heinisch} C. Heinisch and M. Holthaus, Journal of Modern Optics {\bf 63}, 18 (2016).
\bibitem{Ovadyahu} Z. Ovadyahu, Phys. Rev. Lett. {\bf 108}, 156602 (2012).
\bibitem{Iwai} S. Iwai, M. Ono, A. Maeda, H. Matsuzaki, H. Kishida, H. Okamoto, and Y. Tokura, Phys. Rev. Lett. {\bf 91}, 057401 (2003).
\bibitem{Kaiser} S. Kaiser, C. R. Hunt, D. Nicoletti, W. Hu, I Gierz, H. Y. Liu, M. L. Tacon, T. Loew, D. Haug, B. Keimer, and A. Cavalleri, 
Phys. Rev. B {\bf 89}, 184516 (2014).
\bibitem{Oka} T. Oka and H. Aoki, Phys. Rev. B {\bf 79}, 081406 (2009).
\bibitem{Kita} T. Kitagawa, T. Oka, A. Brataas, L. Fu, and E. Demler, Phys. Rev. B {\bf84}, 235108 (2011).
\bibitem{Linder} N. H. Linder, G. Refael, and V. Galistki, Nat. Phys. {\bf7}, 490 (2011).
\bibitem{Wang} Y. H. Wang, H. Steinberg, P. Jarillo-Herrero, and N. Gedik, Science {\bf 342}, 453 (2013).
\bibitem{Hai} W. Hai, C. Lee, G. Chong, and L. Shi, Phys. Rev. E {\bf{66}}, 026202 (2002).
\bibitem{Salmond} G. L. Salmond, C. A. Holmes, and G. J. Milburn, Phys. Rev. A {\bf{65}}, 033623 (2002).
\bibitem{Xie} Q. Xie, W. Hai, and G. Chong, Chaos {\bf{13}}, 801 (2003).

\bibitem{Wright} M. J. Wright, J. A. Pechkis, J. L. Carini, and P. L. Gould, Phys. Rev. A {\bf{74}}, 063402 (2006).
\bibitem{Carini} {\it{Production of Ultracold Molecules with Chirped Nanosecond-Timescale Pulses}}, J. L. Carini, (2006).
\bibitem{Wu} B. Wu and Q. Niu, Phys. Rev. A {\bf{61}}, 023402 (2000).
\bibitem{Lu} X. Luo, Q. Xie, and B. Wu, Phys. Rev. A {\bf{77}}, 053601 (2008).
\bibitem{Har} H. L. Haroutyunyan and G. Nienhuis, Phys. Rev. A {\bf{70}}, 063603 (2004).
\bibitem{Lign} H. Lignier, C. Sias, D. Ciampini, Y. Singh, A. Zenesini, O. Morsch, and E. Arimondo, Phys. Rev. Lett. {\bf{99}}, 220403 (2007).
\bibitem{G} J. Gong, L. M. -Molina, and P. H\"{a}nggi, Phy. Rev. Lett. {\bf{103}}, 133002 (2009).
\bibitem{Eck} A. Eckardt, M. Holthaus, H. Lignier, A. Zenesini, D. Ciampini, O. Morsch, and E. Arimondo, Phys. Rev. A {\bf{79}}, 013611 (2009).
\bibitem{gati} R. Gati and M. K. Oberthaler, J.Phys. B: At. Mol. Opt. Phys. {\bf{40}}, R61 (2007).
\bibitem{bishop} A. R. Bishop,M. G. Forest, D. W. McLaughlin, and E. A. Overman II, Physica D, {\bf 23}, 293-328 (1986).
\bibitem{Oteski} L. Oteski, Y. Duduet, L. Pastur, and P. L. Qu\'{e}r\'{e}, Phys. Rev. E {\bf 92}, 0433020 (2015).
\bibitem{mishra} P. K. Mishra, J. Herault, S. Fauve, and M. K. Verma, Phys. Rev. E {\bf 91}, 053005 (2015).
\bibitem{period double} M. V. Jackobson, Commun. Math. Phys. {\bf 81}, 39-88 (1981).
\bibitem{intermittency} A. Prasad, V. Mehra, and R. Ramaswamy, Phys. Rev. Lett. {\bf79}, 21 (1997).
\bibitem{crisis} C. Grebogi, E. Ott, J. A. Yorke, Physica D, {\bf 7}, 181-200 (1983).
\bibitem{crisisInter} G. Tanaka, M. A. F. Sanjuan, and K. Aihara, Phys. Rev. E, {\bf 71}, 016219 (2005).
\bibitem{RTN} G. J. de Valcarcel, E. Roldan, and R. Vilaseca, Phys. Rev. A  {\bf 45}, R2674 (1992).
\bibitem{H} H. L. Haroutyunyan and G. Nienhuis, Phys. Rev. A {\bf{70}}, 063603 (2004).
\bibitem{W} G. Watanabe and H. M\"{a}kel\"{a}, Phys. Rev. A {\bf{85}}, 053624 (2012).
\end{thebibliography}
\end{document}